\documentclass[aps,pre,preprint,superscriptaddress]{revtex4}

\usepackage{graphicx}

\usepackage{bm}

\usepackage[mathscr]{eucal}

\usepackage{graphics}
\newcommand{ \myvect } [1] { {\bm {#1}} }
\newcommand { \myddot } [2] { { {#1} \cdot {#2} } }

\newcommand { \timestwopi } { { } }

\newcommand { \corrsymbol } { corr }
\newcommand { \uncorrsymbol } { uncorr }
\newcommand { \myotherI } { \mathscr{I} }
\newcommand { \Itimeint } { \mathscr{I} }

\newcommand{ \myop } [1] { {\hat {#1}} }
\newcommand{ \classav } [1] { {\langle {#1} \rangle _R} }

\begin{document}



\title{ Effect of two--particle correlations on x--ray coherent
diffractive imaging studies performed with continuum models}



\author{Zoltan Jurek 
  \footnote{Corresponding author. Email: zoltan.jurek@cfel.de} 
  \footnote{On leave from Research Institute for Solid State Physics 
            and Optics of the Hungarian Academy of Sciences, 
			1525 Budapest P.O.Box 49., Hungary}  }
\affiliation{ 
	Center for Free-Electron Laser Science,
	Deutsches Elektronen-Synchrotron,
	Notkestrasse 85, D-22607 Hamburg, Germany
}

\author{Robert Thiele}
\affiliation{ 
	Center for Free-Electron Laser Science,
	Deutsches Elektronen-Synchrotron,
	Notkestrasse 85, D-22607 Hamburg, Germany
}

\author{Beata Ziaja}
\affiliation{ 
	Center for Free-Electron Laser Science,
	Deutsches Elektronen-Synchrotron,
	Notkestrasse 85, D-22607 Hamburg, Germany
}
\affiliation {
	Institute of Nuclear Physics, Polish Academy of Sciences,
	Radzikowskiego 152, 31-342 Krakow, Poland
}

\author{Robin Santra}
\affiliation{ 
	Center for Free-Electron Laser Science,
	Deutsches Elektronen-Synchrotron,
	Notkestrasse 85, D-22607 Hamburg, Germany
}
\affiliation {
	Department of Physics, University of Hamburg,
	Jungiusstrasse 9, 20355 Hamburg, Germany
}

\date{\today}


\begin{abstract}
Coherent diffraction imaging (CDI) of single molecules at atomic resolution is 
a major goal for the x-ray free electron lasers (XFELs). However, during an imaging  pulse, the fast laser-induced ionization may strongly affect the recorded diffraction pattern of the irradiated sample. The radiation tolerance of the imaged molecule should then be investigated 'a priori' with a dedicated simulation tool. The continuum approach is a powerful tool for modeling the evolution of irradiated large systems consisting of more than a few hundred thousand atoms.  However, this method follows the evolution of average single-particle densities, and the experimentally  recorded intensities reflect the spatial two-particle correlations. The information on these correlations is then inherently not accessible within the continuum approach. In this paper we analyze this limitation of continuum models and discuss the applicability of continuum models for imaging studies.  We propose a formula to calculate  scattered intensities (including both elastic and inelastic scattering) 
from the estimates obtained with a single-particle continuum model. We derive this formula for systems under conditions typical for CDI studies with XFELs. \end{abstract}


\maketitle


\section{\label{section_intro} Introduction}

Coherent diffraction imaging of single molecules at atomic resolution \cite{Neutze2000} is a major goal for the present and forthcoming x-ray free electron lasers (LCLS, SACLA, European XFEL) \cite{LCLS,SACLA-1st-Nature,XFEL}. The high intensity and the extremely short pulse length ($<$\,100\,fs) of the XFEL beam are needed to get a sufficiently strong scattering signal from the sample and to reduce the
effect of radiation damage (in comparison to low fluence imaging experiments, where the damage accumulates over many shots). The sample is destroyed during each measurement (so--called {\it diffract--and--destroy} method \cite{Neutze2000}). A single
diffractive pattern is so noisy that averaging over patterns corresponding to
the same molecular orientation is needed in order to increase the
signal-to-noise ratio \cite{Huldt_2003,Bortel2007,ShneersonActaCryst2008,veit_2009,Bortel2009,fung_2009}. So for a 3D reconstruction, many patterns of the same molecule are needed.

X-ray diffraction gives information on the electron density of the system, so
any changes of the electron density due to the radiation damage influence the
diffraction patterns. There are three main damage components: (i) atoms are 'loosing' bound electrons due to ionization, (ii) ionized atoms may move from their original positions, and (iii) scattering on the increasing free-electron density contributes to the signal. In order to investigate the effect of these damage components, detailed modeling and understanding of ionization dynamics are needed. The continuum approach (Boltzmann \cite{Beata2006}, hydrodynamic \cite{HauRiegePRE2004}) is an efficient way of modeling damage within large systems. However, the presently available continuum methods follow the average single-particle densities, and the recorded diffraction intensities reflect two-particle correlations. The information on two-particle correlations is inherently not accessible from a single-particle continuum approach. 

In this paper we analyze this limitation of continuum models and discuss the applicability of continuum models for imaging studies.  
In order to study this specific problem we do not need to construct a full 'ab initio' model of radiation damage. We analyze this problem with a simplified molecular dynamics model reproducing the conditions typical for  coherent diffraction imaging studies that are known from experiments and from the theoretical  studies of radiation damage \cite{HauRiegePRE2004}. The structure of the paper is as follows. In section II we define the relation between total (elastic and inelastic) diffracted signal and its corresponding estimate from the continuum model. In section III we describe the physical conditions developing within XFEL irradiated samples. In section IV we derive an approximate formula for signal scattered off an XFEL irradiated sample. It is derived from the estimates obtained with single-particle densities. This formula is then tested against numerical simulations in section V. The results obtained are also discussed therein. The Appendix contains information on the simulation details. Finally, in section VI our conclusions are listed. 


\section{\label{section_system} Imaging studies and continuum models}

The state of an imaged sample during a CDI experiment is dynamic, it changes with time as a consequence of the sample irradiation. Stochastically occurring ionizations change the electronic states of the atoms at different times, the density of the released free electrons increases. Therefore the results of a CDI experiment, even when repeated under the same conditions, will vary, due to the stochasticity. We call a unique time evolution that occurs during one CDI experiment a single {\it realization}.

In a single realization the total (elastic and inelastic) x-ray intensity scattered off a system of electrons in state $\vert \Psi \rangle$ at time $t$ is proportional to \cite{JamesBook,Hubbell}:
\begin{eqnarray}
I ({\myvect{q}},t) =
\int \!\! d^{3} r \, d^{3} r' \, \langle \Psi \vert \, \myop{n}({\myvect{r}},t) \, \myop{n}({\myvect{r} \,'},t) \, \vert \Psi \rangle \,
e^{ \timestwopi i \myddot  {\myvect{q}}  { ( {\myvect{r}} - {\myvect{r} \,'} )}  }  =
\langle \Psi \vert \, \myop{n}({\myvect{q}},t) \, \myop{n}^{\star}({\myvect{q} },t) \, \vert \Psi \rangle
\label{eq1}
\end{eqnarray}
where $\myop{n}({\myvect{r}},t)$ and $\myop{n}({\myvect{q}},t)$ are the electron density operator, and its Fourier-transform, respectively:
\begin{eqnarray}
\myop{n}(\myvect{r},t) &=& \sum _{j=1}^{N_{el}} \delta (\myvect{r} - \myop{\myvect{r}}_j(t)) ,
\nonumber \\
\myop{n}(\myvect{q},t) &=& \int \!\! d^3 r \, \myop{n}(\myvect{r},t) \,e ^{ \timestwopi i \myddot {\myvect{q}} {\myvect{r}}  }  =
\sum _{j=1}^{N_{el}} e ^{ \timestwopi i \myddot {\myvect{q}} {\myop{\myvect{r}}_j}(t) }, 
\label{eq:Itimeintdef}
\end{eqnarray}
and the operator $\myop{\myvect{r}}_j(t)$ is the position operator in the Heisenberg picture.

Assuming the coherence time of the pulse to be short compared to the timescales of the processes occurring within the irradiated sample, the time integrated intensity can be approximated as an incoherent sum of the intensities scattered at instantaneous snapshots of the system. The number of photons scattered at the vector ${\myvect{q}}$ during a single realization is then proportional to:
\begin{eqnarray}
\Itimeint ({\myvect{q}}) \equiv \int \!\! dt \, h(t) \, I ({\myvect{q}},t).
\label{eq:Itimeintdef}
\end{eqnarray}
The function $h(t)$ describes the average temporal envelope of the pulse, i. e. the modulus of the pulse amplitude squared, which is ensemble-averaged over XFEL shots. During an imaging experiment a large number of patterns from single shots is collected and summed up, in order to obtain an average pattern. Therefore the average of the recorded signal {\it over the measured realizations (R)} must be formed: 
\begin{equation}
\langle \Itimeint ({\myvect{q}})\rangle _{R} =
\int \!\! dt \, h(t) \int \!\!  d^{3}r \, d^{3} r' \, \langle \myop{n}({\myvect{r}},t) \, \myop{n}({\myvect{r} \,'},t) \rangle _{R} \;
e^{ \timestwopi{} i \myddot {\myvect{q}} { ( {\myvect{r}} - {\myvect{r} \,'} ) } }.
\label{eq4}
\end{equation}
For operators the symbol $\langle ... \rangle _R$ implies the averaging of the expectation value over realizations, i.\ e.\ for an observable $\myop{A}$:
\begin{equation}
 \langle \myop{A} \rangle _R \equiv \sum p_n(t) \langle \Psi _n \vert \myop{A} \vert \Psi _n \rangle ,
\label{eq5}
\end{equation}
where $p_n(t)$ is the probability that the system is in one of the realized states $\vert \Psi _n \rangle$. For a classical quantity $B$ averaging over realization simplifies to:
\begin{equation}
 \langle B \rangle _R \equiv \sum p_n(t) B_n 
\end{equation}
where the quantity, $B$, assumes a value, $B_n$, with a probability, $p_n(t)$, at time $t$.  Here we would like to emphasize that our analysis is not restricted to systems in thermal equilibrium, so that the different realizations are not corresponding merely to the thermal fluctuations of an equilibrated system but may also lead through the non-equilibrium stages of the system evolution. 

The continuum models, used to simulate the evolution of the irradiated systems, describe dynamical properties of electrons and ions, using average single--particle densities $\langle \myop{n}({\myvect{r}},t) \rangle _{R}$. The intensity that one can construct directly from these average densities is:
\begin{equation}
\Itimeint ^{C}({\myvect{q}}) =
\int \!\!  dt \, h(t)  \int \!\! d^{3}r \, d^{3} r' \;
\langle \myop{n}({\myvect{r}},t) \rangle _{R} \: \langle \myop{n}({\myvect{r} \,'},t) \rangle _{R}  \;
e^{ \timestwopi{} i \myddot {\myvect{q}} { ( {\myvect{r}} - {\myvect{r} \,'} )} }
.
\label{eq:Ic}
\end{equation}
The difference between $\langle \Itimeint (\myvect{q}) \rangle _R $ and $\Itimeint ^C(\myvect{q})$ depends on two--particle correlations during individual realizations. 


\section{\label{section_system} Evolution of x-ray irradiated biomolecules under the conditions typical for imaging studies}

Detailed simulations of radiation damage
\cite{Neutze2000,jurekEPJD2004,HauRiegePRE2004,dresdenbiol} reveal the typical ionization dynamics within an x-ray irradiated biological sample during an imaging experiment. Biological samples are typically built up from light elements like C, N, O, H. Only a few atoms of heavy elements are present, e.\ g.\ Cl, S, Fe. The primary events that initiate radiation damage are the photoionization events \cite{SantraJPBtutorial}. Photoelectrons released by x-rays from light elements have high energy ($\sim10\,$keV). They can leave the irradiated system within a few fs, so that the net charge of the system starts to increase. During inner-shell photoionization, core holes are created. These holes are then filled by electrons from the outer shells within a few tens of fs. For light elements Auger decay is dominant \cite{SantraJPBtutorial}. Auger electrons have lower energies ($\sim$ few hundred eVs), and they can be trapped within the charged sample. These quasi-free electrons cause further ionizations that end up in secondary electron cascading \cite{ZiajaPRB2001,ZiajaPRB2002}.  

It is energetically preferred for the quasi-free electrons to move towards the
center of the sample, forming a net-neutral core with ions. This leads to the
formation of a positively charged outer shell of ions around the neutral core.
For samples large enough and sufficiently highly ionized, even high-energy photoelectrons can be trapped in the core, enhancing the temperature of the quasi-free electrons.

Conditions of the electron plasma can be characterized by comparing the average potential energy of the electrostatic interaction between the electrons, $E_{Coulomb}$, to their temperature, $k_B T$
\cite{GlenzerRedmer2009,MurilloPhysRep1998}:
\begin{equation}
\Gamma = { {E_{Coulomb}} \over {k_B T} }
= {{1} \over {4 \pi \epsilon _{0} }} \, { {e^2} \over {k_{B}T} } \, \left ({{4 \pi n} \over 3} \right ) ^ {{{1} \over {3}}} 
,
\end{equation}
where $n$ is the density of the electrons. In case of $\Gamma \gtrsim 1$, Coulomb interaction dominates the electron dynamics and the plasma is strongly coupled. If $\Gamma \lesssim 1$, the Coulomb interaction is suppressed, and 
the plasma is weakly coupled.  

Another parameter, the ratio of the electron temperature and the
Fermi energy, $\Theta$, gives information on the quantum effects within the system \cite{GlenzerRedmer2009,MurilloPhysRep1998}: 
\begin{equation}
\Theta = { {k_{B}T} \over {E_{F}} } = 
{ {k_B T} \over { { {\hbar ^2} \over {2m_e} } (3 \pi ^2 n)^{2/3} } }
.
\end{equation}
At electron temperatures higher than the Fermi energy ($\Theta \gtrsim 1$),
a classical description is adequate, while at $\Theta \lesssim 1$, 
a quantum mechanical treatment is needed.

According to the aforementioned simulations \cite{HauRiegePRE2004}, the typical
electron density at the center of a biomolecule is around 10$^{23}$\,cm$^{-3}$, the electron temperature is $\sim 20\,$eV, when photoelectrons are not trapped, and $\ge100\,$eV when photoelectrons are trapped. For the values of temperature, $T = 20$\,eV and $T = 100$\,eV, the plasma parameters are $\Gamma \approx 0.5$, $\Theta \approx 2.5$, and $\Gamma \approx 0.1$, $\Theta \approx 12.5$, respectively, so that the plasma of quasi-free electrons can be treated as classical and ideal.

It should be noted that ions do not form a plasma during an imaging experiment.
As the imaging pulses are short ($\sim$tens of femtoseconds) compared to electron-ion equilibration time ($\sim$few ps), ions remain 'cold' during the pulse. Unscreened ions from the outer shell can move radially, due to the repulsive Coulomb forces. However, if the pulse is not longer than several fs, those unscreened ions will also not relocate during the pulse \cite{HauRiegePRE2004,jurekEPJD2008}. As a result, during the imaging pulse, the initially neutral system evolves into a two-regime system consisting of 'frozen' ions embedded in a thermalized free-electron plasma. 


\section{\label{section_theory} Improved description of diffractive signal obtained from  single-particle continuum model for XFEL irradiated sample}

In this section we investigate in detail how the scattered signal averaged over the realizations reflects the changing electron density within the irradiated sample. We follow the approach proposed by Chihara in Ref.\ \cite{ChiharaJPhysMet1987}, and further applied in many works investigating x-ray scattering on plasmas, e.\ g.\ 
\cite{ChiharaJPhys2000,NardiPRA1991,GregoriPRE2003,HollHEDP2007}. However, in
our case we will apply it to a system that can be far from thermal equilibrium at its initial evolution stages. This is in contrast to the standard plasma applications that usually consider systems in thermal equilibrium.

Following \cite{ChiharaJPhysMet1987}, we separate the total electron density operator, $\myop{n}({\myvect{r}},t)$, into the the bound electron component,
$\myop{n}_{b}({\myvect{r}},t)$, and the unbound electron component. In general, one could develop a model for describing the dynamics of all electrons, also those ones that escape from the sample. However, the currently available continuum models \cite{HauRiegePRE2004,Beata2006} follow the dynamics of particles only within a restricted volume. Therefore we separate the unbound electron density operator into two components. One component refers to the electrons that escaped from the sample, $\myop{n}_{e}({\myvect{r}},t)$, and the other one to the electrons trapped inside the sample, $\myop{n}_{t}({\myvect{r}},t)$.  
From Eqs.~(\ref{eq1}), (\ref{eq5}) we then obtain:
\begin{eqnarray}
\langle I( {\myvect{q}}, t ) \rangle _{R} & = &
\left \langle \left \vert 
	\int \left ( \myop{n}_b({\myvect{r}},t) + \myop{n}_{t}({\myvect{r}},t) + \myop{n}_{e}({\myvect{r}},t) \right ) \,
	e^{ \timestwopi{} i \myddot {\myvect{q}} {\myvect{r}} } \, d^{3}r 
\right \vert ^{2} \right \rangle  _{R}
\nonumber \\
& = & \left \langle \left \vert \myop{n}_b({\myvect{q} },t) \right \vert ^{2} \right \rangle _{R}  +  
\left \langle \left \vert \myop{n}_{t}({\myvect{q}},t ) \right \vert ^{2} \right \rangle _{R}  +  
\left \langle \left \vert \myop{n}_e({\myvect{q}},t ) \right \vert ^{2} \right \rangle _{R}    
\nonumber \\
& & + 2 \, Re \, \left [  \,
 \left \langle  \myop{n}_b({\myvect{q} } ,t) \, \myop{n}_{t}^{\star} ({\myvect{q}},t ) \right \rangle _{R} +
 \left \langle \myop{n}_b({\myvect{q} },t ) \, \myop{n}_{e}^{\star} ({\myvect{q}},t ) \right \rangle _{R} + 
 \left \langle \myop{n}_{t}({\myvect{q} } ,t) \, \myop{n}_{e}^{\star} ({\myvect{q}},t ) \right \rangle _{R} \, \right ] 
\nonumber \\
& = & \langle I_{bb}(\myvect{q},t) \rangle _{R} + \langle I_{tt}({\myvect{q} },t ) \rangle _{R} + \langle I_{ee}({\myvect{q} },t ) \rangle _{R} 
\nonumber \\
& & + 2 \, Re \, \left [ \langle I_{bt}({\myvect{q} },t ) \rangle _{R} + \langle I_{be}({\myvect{q} },t ) \rangle _{R} + \langle I_{te}({\myvect{q} },t ) \rangle _{R} \, \right ],
\end{eqnarray}
where $I_{ij}(\myvect{q},t) \equiv\langle \Psi \vert \, \myop{n}_i({\myvect{q}},t) \, \myop{n}_j^{\star}({\myvect{q} },t) \, \vert \Psi \rangle$ for $i,j=b,t,e$. First we analyze the effect of the diagonal terms, $ \langle I_{bb}(\myvect{q},t) \rangle _{R}$, $ \langle I_{tt}(\myvect{q},t) \rangle _{R}$,
$ \langle I_{ee}(\myvect{q},t) \rangle _{R}$, and later the effect of the cross terms, $ \langle I_{bt}(\myvect{q},t) \rangle _{R}$, $ \langle I_{be}(\myvect{q},t) \rangle _{R}$,
$ \langle I_{te}(\myvect{q},t) \rangle _{R}$.

\subsection{\label{subsection_bound} Bound electrons }

We can express the bound electron density operator as a convolution of atomic electron densities, $\myop{n}_{b,j}({\myvect{r}},t)$ and of the atomic position operators, $\delta ({\myvect{r}}-\myop{\myvect{R}}_j(t))$ \cite{DensConvol}:
\begin{eqnarray}
\myop{n}_b({\myvect{r} },t) = \sum _{j=1} ^{N_a} \int d^3r' \,\myop{n}_{b,j} ( {\myvect{r}\,'},t)
\delta( {\myvect{r} } - {\myvect{r} \,'}- \myop{{\myvect{R}}}_j(t) ).
\end{eqnarray}
Its Fourier transform then reads:
\begin{eqnarray}
\myop{n}_b({\myvect{q}},t) = 
\int \myop{n}_b({\myvect{r}},t) \, e ^ { \timestwopi i \myddot {\myvect{q}} {\myvect{r}}   } \, d^3 r = 
\sum _{j=1} ^{N_a} \myop{n}_{b,j} ( {\myvect{q}},t) \, 
e ^ { \timestwopi i \myddot {\myvect{q}} {\myop{{\myvect{R}}}_j(t)}   }, 
\end{eqnarray}
where $N_a$ is the total number of atoms, $\myop{{\myvect{R}}}_j(t)$ is the position operator of atom $j$, and $\myop{n}_{b,j} ( {\myvect{q}},t)$ is the Fourier transform of the density $\myop{n}_{b,j} ( {\myvect{r}},t)$. The scattered intensity averaged over realizations is:
\begin{eqnarray}
\label{eq:Ibound_def}
 \langle I_{bb}( {\myvect{q}} ,t ) \rangle _{R} = 
\left \langle \left \vert \myop{n}_{b}( {\myvect{q}},t) \right \vert ^2 \right \rangle _R =
\sum _{j,k=1} ^{N_a} 
	\left \langle \, \myop{n}_{b,j}( {\myvect{q}},t)\,
	\myop{n}_{b,k}^{\star} ( {\myvect{q}},t) \, 
	e ^ { \timestwopi i \myddot {\myvect{q}} { ( \myop{{\myvect{R}}}_j(t) -\myop{{\myvect{R}}}_k(t) ) } } \, \right \rangle _R
.
\end{eqnarray}
Radiation damage can affect the intensity pattern in two ways: (i) form factors
of atoms can decrease due to the progressing ionization, (ii) positions of ions can change due to the Coulomb interaction within the ionized system. The latter effect can be overcome by applying pulses short enough ($\lesssim 10\,$fs
\cite{jurekEPJD2008}), which are available at XFELs \cite{shortXFEL}.
However, the reduction of the form factors cannot be eliminated.

Assuming static atomic positions,
$\langle \myop{\myvect{R}}_j(t) \rangle_R\equiv \myvect{R}_j$ in all realizations, and no correlation between ionization events of different atoms, so that 
$\langle \myop{n}_{b,j}( {\myvect{q} },t) \myop{n}_{b,k}^{\star} ( {\myvect{q} },t ) \rangle _R =
\langle \myop{n}_{b,j}( {\myvect{q} },t) \rangle _R \, \langle \myop{n}_{b,k}^{\star} ( {\myvect{q} },t)
\rangle _R$ for $j \neq k$, the corresponding averaged intensity is:
\begin{eqnarray}
\label{eq:bound_final}
 \langle I_{bb}^{\uncorrsymbol}( {\myvect{q}} ,t ) \rangle _{R}  & = & 
\sum _{j=1} ^{N_a} \left \langle \left \vert \myop{n}_{b,j}( {\myvect{q} },t)  \right \vert ^2 \right \rangle _R +
\sum _{j \neq k=1} ^{N_a} 
	\left \langle \myop{n}_{b,j}( {\myvect{q} },t)  \right \rangle _R \, \left \langle  \myop{n}_{b,k}^{\star} ( {\myvect{q} },t) \right \rangle _R
	\; e ^ { \timestwopi i \myddot {\myvect{q}} { ( {\myvect{R}}_j -{\myvect{R}}_k ) } }  
\\
& = & \sum _{j=1} ^{N_a} \classav {I^{inel}_j}   +   
      \sum _{j=1} ^{N_a} \left ( \, \classav {  \left \vert f_{j}( {\myvect{q} },t )  \right \vert ^2 } -
				\left \vert \, \classav {  f_{j}( {\myvect{q} },t )  } \, \right \vert ^2 \, \right ) +
\left \vert \, \classav{ \sum _{j = 1} ^{N_a} f_{j} ( {\myvect{q} },t ) 
	\; e ^ { \timestwopi i \myddot {\myvect{q}} {\myvect{R}}_j }  } \, \right \vert ^2
\nonumber \\
& = & \sum _{j=1}^{N_a} { \classav{N_{b,j}(t)} \over {Z_j} } {I}^{inel}_{atomic,Z_j}   +
      \sum _{j=1}^{N_a} \left ( \, \classav{ \left \vert f_{j}( {\myvect{q} },t )  \right \vert ^2 } -
				\left \vert \, \classav{  f_{j}( {\myvect{q} },t )  } \, \right \vert ^2 \, \right ) +
\left \vert \left \langle  n_{b} ( {\myvect{q} },t ) \right \rangle _R \right \vert ^2\nonumber,
\end{eqnarray}
where $f_{j}( {\myvect{q} },t ) \equiv \,\left \langle \Psi \mid \myop{n}_{b,j}( {\myvect{q} },t)  \mid \Psi \right \rangle$ is the atomic form factor, $ n_{b} ( {\myvect{q} },t )\equiv \sum _{j = 1} ^{N_a} f_{j} ( {\myvect{q} },t ) \; e ^ { \timestwopi i \myddot {\myvect{q}} {\myvect{R}}_j }$ is the classical bound electron density. Deriving the second row in Eq.\ (\ref{eq:bound_final}), we used the following relation:
\begin{eqnarray}
\label{eq:bound_final_x}
\left \langle \left \vert \myop{n}_{b,j}( {\myvect{q} } )  \right \vert ^2 \right \rangle _R & = &
\sum _n p_n(t)  \left ( \, 
  \langle \Psi _n \vert \myop{n}_{b,j}( {\myvect{q} } ) \vert \Psi _n \rangle
  \langle \Psi _n \vert \myop{n}_{b,j}^{\star}( {\myvect{q} } ) \vert \Psi _n \rangle \right.
\nonumber \\
  & & \, + \, \left. \langle \Psi _n \vert \myop{n}_{b,j}( {\myvect{q} } ) \;  ( 1 -\vert \Psi _n \rangle
  \langle  \Psi _n \vert ) \,  \myop{n}_{b,j}^{\star}( {\myvect{q} } ) \vert \Psi _n \rangle  
\right )
\nonumber \\
& = & \classav{ \left \vert f_{j}( {\myvect{q} },t )  \right \vert ^2 } + \classav {I^{inel}_j}.
\label{replace}
\end{eqnarray}
If the binding energy of a bound electron is much less than the energy of the incoming photon, the intensity ${I}^{inel}_j$ scattered inelastically on the atom/ion, $j$, can be approximated as being proportional to the number of its bound electrons, $N_{b,j}(t)$, multiplied by the intensity,  ${I}^{inel}_{atomic,Z_j}$, scattered inelastically from the neutral element of atomic number, $Z_j$:
\begin{eqnarray}
\label{eq:bound_final_x}
\classav {I^{inel}_j} = { \classav{N_{b,j}(t)} \over {Z_j} } {I}^{inel}_{atomic,Z_j} \,.
\end{eqnarray}  

According to Eq.~(\ref{eq:bound_final}) (last row), the total averaged scattered intensity from bound electrons is then the sum of: (i) the inelastic incoherent scattering component, $\sum _{j=1} ^{N_a} \classav {I^{inel}_j}$, (ii) elastic incoherent scattering component containing the variance of the atomic form factors of the individual ions, and (iii) the density, $\left \vert \left \langle  n_{b} ( {\myvect{q} },t ) \right \rangle _R \right \vert ^2$  that describes coherent elastic scattering. Similar terms appear in the final formula of Ref. \cite{ChiharaJPhysMet1987}, except for that term which is a consequence of the externally driven ionization dynamics.

\subsection{\label{subsection:trapped} Trapped electrons }

We investigate now the contribution of trapped electrons to the imaging pattern. Under the conditions typical for CDI studies electrons can be treated as classical particles. In what follows we will then replace the electron-density-operator expectation values, $\langle \Psi \vert \,\myop{n}({\myvect{r}},t) \vert \Psi  \,\rangle$, with their corresponding classical equivalents,
$n ({\myvect{r}},t)$. The classical electron density and its Fourier transform, defined for $N_t$ trapped electrons, located at positions ${\myvect{r}}_{j}(t)$ at time $t$, then are:
\begin{eqnarray} 
n_t({\myvect{r}},t)  &=&
\sum _{j=1}^{N_t} \delta ( {\myvect{r}} - {\myvect{r}}_{j}(t) )
,\nonumber\\
n_t({\myvect{q}},t)  &=&
\int \!\! d^{3} r \; n_t ({\myvect{r}},t) \; e^{ \timestwopi i \myddot {\myvect{q}} {\myvect{r}} } = 
\sum _{j=1} ^{N_t} e^{ \timestwopi i \myddot {\myvect{q}} {\myvect{r}}_{j}(t) } 
,
\end{eqnarray}
and the realization-averaged signal scattered from trapped electrons is proportional to:
\begin{equation}
\langle I_{tt}({\myvect{q}},t) \rangle _{R}  =
N_t + \sum _{j \neq k =1}^{N_t} \left \langle e^{ \timestwopi i \myddot {\myvect{q}} { ( {\myvect{r}}_{j}(t)- {\myvect{r}}_{k}(t) ) } } \right \rangle _{R}
.
\label{eq:Iav}
\end{equation}
Using the two-particle distribution function \cite{HansenMcD}, one can rewrite it further as:
\begin{eqnarray}
\label{eq:trapped_final_corr}
\langle I_{tt}(\myvect{q},t) \rangle _R  \, & = & \,  
N_t + 
{ {N_t - 1} \over {N_t} }  \int\!\! d^3r d^3r' 
   \left \langle n_t({\myvect{r}},t) \right \rangle _{R}
   \left \langle n_t({\myvect{r}'},t) \right \rangle _{R} 
   g(\myvect{r},\myvect{r}',t) 
   e^{ \timestwopi i \myddot {\myvect{q}} { (\myvect{r} - \myvect{r}') } }  
\nonumber \\
& = & N_t + { {N_t - 1} \over {N_t} } \left \vert \left \langle n_t({\myvect{q}},t) \right \rangle _{R} \right \vert ^2
\nonumber \\
& & \,\,\, + \,{ {N_t - 1} \over {N_t} }  \int\!\! d^3r d^3r'  \left \langle n_t({\myvect{r}},t) \right \rangle _{R}
   \left \langle n_t({\myvect{r}'},t) \right \rangle _{R} ( g(\myvect{r},\myvect{r}',t) - 1 ) 
   e^{ \timestwopi i \myddot {\myvect{q}} { (\myvect{r} - \myvect{r}') } } ,
\label{itt}
\end{eqnarray}
where $g(\myvect{r},\myvect{r}',t)$ is the two-particle distribution function
at time $t$. The time-dependence of $g(\myvect{r},\myvect{r}',t)$ is imposed
by the non-equilibrium evolution of the considered system. 

In the second row of Eq.\ (\ref{itt}) the first term is a $q$-independent shift given by the number of trapped electrons, $N_t$. The second term depends on the average electron density. The third term contains the statistical information about the two-particle correlations. This information is not accesible within single-particle continuum approach, as considered here. The first term and the third one correspond to the incoherent scattering, whereas the second one corresponds to coherent scattering on free electrons as in Chihara's formula in Ref.\ \cite{ChiharaJPhysMet1987}.

Neglecting all correlations in (\ref{itt}), we arrive at the following estimate:
\begin{eqnarray}
\label{eq:trapped_final}
\langle I_{tt}^{uncorr} (\myvect{q},t) \rangle _R  \, =  N_t + { {N_t - 1} \over {N_t} } \left \vert \left \langle n_t({\myvect{q}},t) \right \rangle _{R} \right \vert ^2.
\label{ittu} 
\end{eqnarray}

In the above derivations we assumed that the number of trapped electrons,
$N_t$, is identical for all realizations at any time $t$. As the $N_t$
distribution results from a stochastic process, this assumption is generally
not valid. In the general case, we should replace $N_t$ with its average over
realizations at time $t$, $\classav { N_t(t) }$. This is an accurate
approximation if the number of particles in individual realizations is
large. For a small number of particles this may introduce a large relative error. However, in our case, the contribution of trapped electrons to the overall scattered intensity is small, so in any case it should not introduce a
significant error.

We note that the approximation, Eq.\ (\ref{ittu}), could be further improved by taking into account correlations. In general, especially in case of short pulses, the electrons may not reach thermal equilibrium during the pulse. This would require a time-dependent parametrization of $g(\myvect{r},\myvect{r}',t)$. If the pulse is long enough compared to the electron thermalization timescale, the two-particle distribution function for electrons could be approximated by the equilibrium one. 

\subsection{\label{subsection:escaped} Escaped electrons }

Energetic photoelectrons leave the sample soon after the exposure to the x-ray
pulse starts.  However, they may stay within the focus of the beam during the
imaging pulse and also contribute to the diffractive patterns.

In order to describe their contribution to the intensity, the same approximate formula as for the trapped electrons, Eq.\,(\ref{eq:trapped_final}), can be applied:
\begin{equation}
 \langle I_{ee}^{\uncorrsymbol}({\myvect{q}},t) \rangle _{R}  =
N_e + {{N_e-1} \over {N_e}} \left \vert \left \langle n_e({\myvect{q}},t) \right \rangle _{R} \right \vert ^2
.
\label{eq:escaped_pre}
\end{equation}
In case of escaped electrons, their density is dilute, e.g., the distances
between escaped electrons are on average large. Their contribution will then
show up mainly at small values of $q$, limited by the inverse of the size of the escaped electron cloud. We explain here that the relevant $q$ range for imaging studies is defined by $1/L < q/2\pi < 2/d$, where $L$ is the size of the object (here the size of the electron cloud), and $d$ is the desired resolution. 
Therefore, if the continuum model does not provide information on $\langle n_e({\myvect{q}},t) \rangle _R$, we can approximate the effect of the escaped electrons as:
\begin{equation}
 \langle I_{ee}^{\uncorrsymbol}({\myvect{q}},t) \rangle _{R} \cong N_e,
\label{eq:escaped_final}
\end{equation}
which is accurate enough in the high $q$ regime, limited by the inverse resolution, $2/d$.  

\subsection{\label{subsection:cross} Cross terms }

The escaping high-energy electrons are typically located at large distances
from other particles, so the potential energy of their individual interaction
with other particles is much smaller than their kinetic energy. Therefore, the
Coulomb correlations between bound electrons and escaped electrons, and between trapped electrons and escaped electrons, can be neglected, and the respective cross terms factorize to:
\begin{eqnarray}
 \langle I_{be}( {\myvect{q}}, t ) \rangle _{R}  & = &  
\langle n_b({\myvect{q} },t ) \, n_{e}^{\star} ({\myvect{q}} ,t) \rangle _R =
\langle n_b({\myvect{q} },t ) \rangle _R \langle  \, n_{e}^{\star} ({\myvect{q}} ,t) \rangle _R, 
\nonumber \\
 \langle I_{te}( {\myvect{q}}, t ) \rangle _{R}  & = &  
\langle n_{t}({\myvect{q} },t ) \, n_{e}^{\star} ({\myvect{q}} ,t) \rangle _R  =
\langle n_t({\myvect{q} },t ) \rangle _R \, \langle  n_{e}^{\star} ({\myvect{q}} ,t) \rangle _R
.
\label{eq:cross_escaped}
\end{eqnarray}
Following our discussion from the previous subsection, we can neglect these
terms in the considered high $q$ regime.

The correlation between bound and trapped electrons, originating from the same
initial process of ionization, strongly decreases with time as the number of
particles increases and the particle cloud evolves. However, there are still
Coulomb correlations between these particles. If we compare the potential energy of the electrons trapped in the field of the ions to their kinetic energy at the typical simulation parameters \cite{HauRiegePRE2004}, we obtain:
\begin{eqnarray}
\Gamma_{ei} & \approx & 0.93 ~~~~ \text{at} ~~~~ T = 20 \, \text{eV}
,
\nonumber \\
\Gamma_{ei} & \approx & 0.18 ~~~~ \text{at} ~~~~  T = 100 \, \text{eV}
.
\end{eqnarray}
If the electron-ion system is weakly coupled ($\Gamma_{ei}<1$), we can then neglect the correlations between the different density terms, and the cross term can be factorized as:
\begin{eqnarray}
 \langle I_{bt}(\myvect{q},t) \rangle _{R} =  
\langle n_b({\myvect{q} },t ) \, n_{t}^{\star} ({\myvect{q}} ,t) \rangle =
 \langle n_b({\myvect{q} },t ) \rangle _R \, \langle  n_{t}^{\star} ({\myvect{q}} ,t) \rangle _R.
\label{eq:cross_trapped}
\end{eqnarray}
Again, the cross terms then depend solely on average densities, which are known from the continuum model. A similar cross term appears in Chihara's formula in Ref.\ \cite{ChiharaJPhysMet1987}.

\subsection{\label{subsection:theory_summary} Intensities from continuum model }

Here we collect the terms that we derived in the previous subsections. The estimate for the total scattered signal in the uncorrelated case in the region of high $q$ values is:
\begin{eqnarray}
 \langle I^{\uncorrsymbol}( {\myvect{q}}, t ) \rangle _{R} & = & 
\sum _{j=1}^{N_a} { \classav{N_{b,j}(t)} \over {Z_j} } {I}^{inel}_{atomic,Z_j}   +
\sum _{j=1} ^{N_a} \left ( \, \left \langle \left \vert f_{j}( {\myvect{q} },t )  \right \vert ^2 \right \rangle _R -
				\left \vert \left \langle  f_{j}( {\myvect{q} },t )  \right \rangle _R \right \vert ^2 \, \right ) +
\left \vert \left \langle  n_{b} ( {\myvect{q} },t ) \right \rangle _R \right \vert ^2  
\nonumber \\
& & + \, N_t + { {N_t -1} \over {N_t} } \left \vert \left \langle n_t({\myvect{q}},t) \right \rangle _{R} \right \vert ^2 \, + \, N_e 
+ \, 2 \, Re \, \left [ \, \left \langle n_b({\myvect{q} },t ) \right \rangle _R \, \left \langle  n_{t}^{\star} ({\myvect{q}} ,t) \right \rangle _R  \, \right ]
\nonumber \\
& = & I^C({\myvect{q} },t) + 
\sum _{j=1} ^{N_a} \left ( \, \left \langle \left \vert f_{j}( {\myvect{q} },t )  \right \vert ^2 \right \rangle _R -
				\left \vert \left \langle  f_{j}( {\myvect{q} },t )  \right \rangle _R \right \vert ^2 \, \right ) 
\nonumber \\
& &
+ \, \sum _{j=1}^{N_a} { \classav{N_{b,j}(t)} \over {Z_j} } {I}^{inel}_{atomic,Z_j}   
- {{1} \over {N_t}} \, I^C_t({\myvect{q} },t)  \, + \, N_t \, + \, N_e
\, ,
\label{eq:final}
\end{eqnarray}
where $I^C$ is the intensity obtained from the total average electron density within the continuum model (cf.\ Eq.\ (\ref{eq:Ic})):
\begin{eqnarray}
I^C({\myvect{q} },t) & = & \left \vert \left \langle n_b({\myvect{q} },t ) \right \rangle _R \, + \left \langle n_t({\myvect{q} },t ) \right \rangle _R \right \vert ^2
\nonumber \\
& = & I^C_b({\myvect{q} },t) \, + \, I^C_t({\myvect{q} },t) \, 
+ \, 2 \, Re \, \left[ \, \left \langle n_b({\myvect{q} },t ) \right \rangle _R \, \left \langle  n_{t}^{\star} ({\myvect{q}} ,t) \right \rangle _R  \, \right],
\end{eqnarray}
and $I^C_{b,t}({\myvect{q} },t)\equiv\left \vert \left \langle n_{b,t}({\myvect{q} },t ) \right \rangle _R \, \right \vert ^2$. If the correlations within the electronic system are known, Eq.\ (\ref{eq:final}) can be improved by including Eq.\ (\ref{itt}). Otherwise, the uncorrelated approximation, Eq.\ (\ref{ittu}), should be used.

\section{\label{section:numerical} Comparison to numerical simulations}

In order to investigate the applicability regime of Eq.\ (\ref{eq:final}),
we performed dedicated molecular dynamics (MD) simulations of electron-ion systems at the conditions corresponding to those that develop in the center of irradiated biological samples during an imaging experiment.
For technical details please see the Appendix. 

Using MD simulations one can follow individual realizations of the dynamics
within an irradiated system. This enables a direct comparison of the time-integrated realization-averaged diffraction signal to its estimate, Eq.\  (\ref{eq:final}), derived from single-particle densities. It corresponds to the estimate that could be obtained from a continuum model. From the results of MD simulations we calculated: (i) the average intensity from different realizations, $\langle I({\myvect{q}},t)\rangle _{R}$ (Eqs.\ (\ref{eq:Itimeintdef}),(\ref{eq4})), (ii) the uncorrelated intensity, $ \langle I^{\uncorrsymbol}({\myvect{q}},t) \rangle _{R}$ (Eq.\ (\ref{eq:final})), and (iii) the correlated intensity, $ \langle I^{\corrsymbol}({\myvect{q}},t) \rangle _{R}$ (Eq.\ (\ref{itt}), for electrons only). We integrated them over time to obtain $\langle \Itimeint(\myvect{q}) \rangle _{R}$, $ \langle \Itimeint ^{\uncorrsymbol}(\myvect{q}) \rangle _{R}$, 
and $ \langle \Itimeint ^{\corrsymbol}(\myvect{q}) \rangle _{R}$ respectively. 
We also calculated $\Itimeint^C(\myvect{q})$, according to Eq.~(\ref{eq:Ic}).
We investigated the following cases.

\begin{itemize}

\item [$\bullet$] { \bf Scattering from bound electrons only }\\
We modelled a cluster consisting of $100$ initially neutral carbon atoms. The
atoms were located randomly within a sphere of radius $R_0=7.1\,$\AA. Their density was $1/15$\,\AA$^{-3}$, which corresponds to a typical protein density. The nearest neighbor distances were always larger than 1.5\,\AA. The atoms were not moving. The time dependence of the average number of bound electrons per ion was approximated by an exponential function  in such a way that the average number of bound electrons per atom decreased from 6 to 4 during the simulation time. 

\item [$\bullet$] { \bf Scattering from trapped electrons only }\\
Electrons were treated as classical particles with regularized Coulomb
interaction between them. The granular ions were not included. However, an
average positive field (radial harmonic potential) was added to keep the
electrons together. The strength of the harmonic potential was determined by
the average ion density and the average ion charge. The electron dynamics was
followed by integrating the equations of motion of the electrons. The parameters we used corresponded to the results of the damage modeling for biosamples \cite{HauRiegePRE2004}: $n_{atom}=1/15\,${\AA}$^{-3}$ (average atomic density), $Q_{ion} = 2$ (average ion charge). The number of electrons was $N=200$. The electrons filled a sphere of $R_0=7.1\,$\AA\ radius at the density of $2 \cdot n_{atom}$. We followed 100 realizations of the system during $T_{pulse}=10$\,fs, from which we calculated the realization average. We performed two sets of calculations, one with the electron temperature of 20\,eV another one with $T_e=$ 100\,eV, which corresponded to the cases when photoelectrons could escape or were trapped within the system, respectively. 

\item [$\bullet$] { \bf Scattering from bound and trapped electrons } \\
This case was just the merging of the above two cases, with the same values of the simulation parameters. A regularized Coulomb potential described the interaction between granular ions and electrons. The external harmonic potential was not applied here.

\end{itemize}

\noindent
Below we discuss the simulation results that we have obtained.

\subsection{\label{subsection:bound} Bound electrons }

\begin{figure}
\includegraphics[width = 10cm]{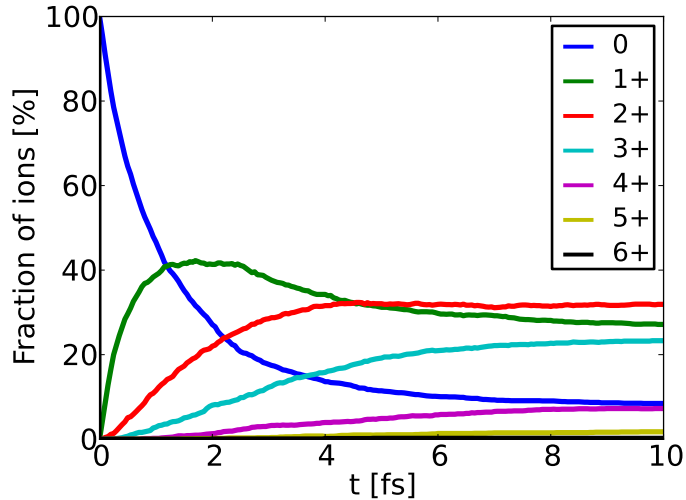} 
\caption{\label{fig:Ifraction} 
(Color online) Average ion-charge-state populations within the imaged system as a function of time.
}
\end{figure}

The scattered intensity depends on stochastic properties of the
ionization process. Fig.\ \ref{fig:Ifraction} shows the fraction of ions with
different charges created within the sample. Although we used a simplified model for describing the ionization (see Appendix), the trends in the charge distributions are in agreement with Ref.~\cite{HauRiegePRE2004}.

\begin{figure}
\includegraphics[width = 10cm]{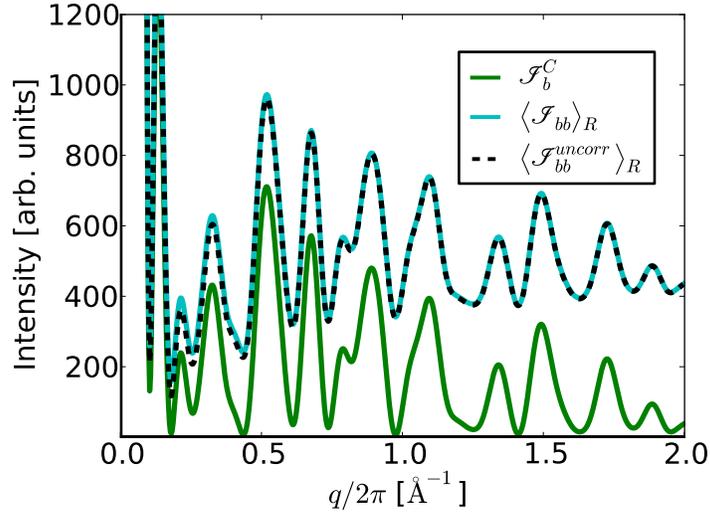} 
\caption{\label{fig:I} 
(Color online) Time integrated intensity from diffractive scattering only by bound electrons as a function of the momentum transfer, $q$, along a random axis in reciprocal space. We show  (i) the intensity constructed from average single particle densities $\myotherI_b^C$, 
(ii) the scattered intensity $\langle \myotherI_{bb}\rangle _{R}$, and (iii) the estimated uncorrelated intensity $\langle \myotherI_{bb}^{\uncorrsymbol}\rangle _{R}$. 
}
\end{figure}

In Fig.~\ref{fig:I} we show the intensity distribution for scattering from bound electrons along a randomly chosen axis in reciprocal space. The $ \langle \myotherI_{bb}^{\uncorrsymbol}(\myvect{q}) \rangle _{R}$
gives a good estimate of $ \langle \myotherI_{bb}(\myvect{q}) \rangle _{R}$. There is a disagreement between $\myotherI^C_b(\myvect{q})$ and $ \langle \myotherI_{bb}(\myvect{q}) \rangle _{R}$, as there is no inelastic scattering included in $\myotherI^C_b(\myvect{q})$. The shift between these two intensities is proportional to the number of bound electrons at high $q$. 

Therefore, one should rather use here $ \langle \myotherI_{bb}^{\uncorrsymbol}(\myvect{q}) \rangle _{R}$  as an estimate of the scattered intensity from bound electrons.

\subsection{\label{subsection:disc_trapped} Trapped electrons }

\begin{figure}
\includegraphics[width = 10cm]{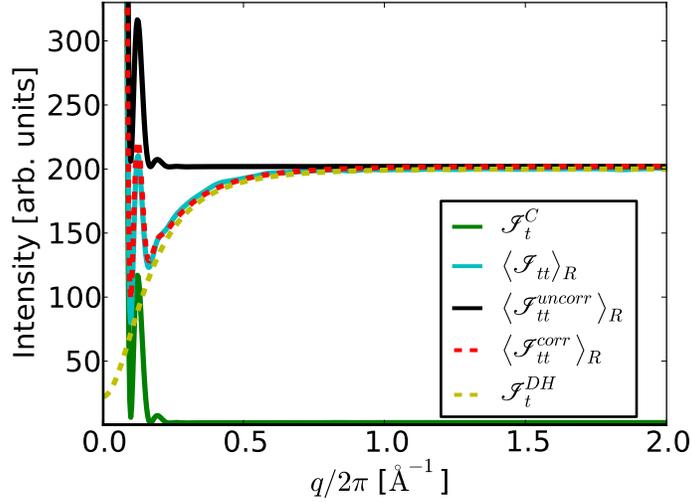} 
\caption{\label{fig:E1h} 
(Color online) Time integrated intensity from diffractive scattering by trapped electrons as a function of the momentum transfer, $q$. The electron temperature was set to 20\,eV, and an external harmonic potential was applied. We show: (i) the intensity constructed from average single particle densities $\myotherI_t^C$, (ii) the scattered intensity $\langle \myotherI_{tt}\rangle _{R}$, (iii) the estimated uncorrelated intensity $ \langle \myotherI_{tt}^{\uncorrsymbol}\rangle _{R}$, 
(iv) the estimated correlated intensity $\langle \myotherI_{tt}^{\corrsymbol}\rangle _{R}$,
and (v) the intensity obtained from the theory of homogeneous, infinite-size, weakly coupled plasmas $\myotherI_{t}^{DH}$. 
}

\end{figure}

\begin{figure}
\includegraphics[width = 10cm]{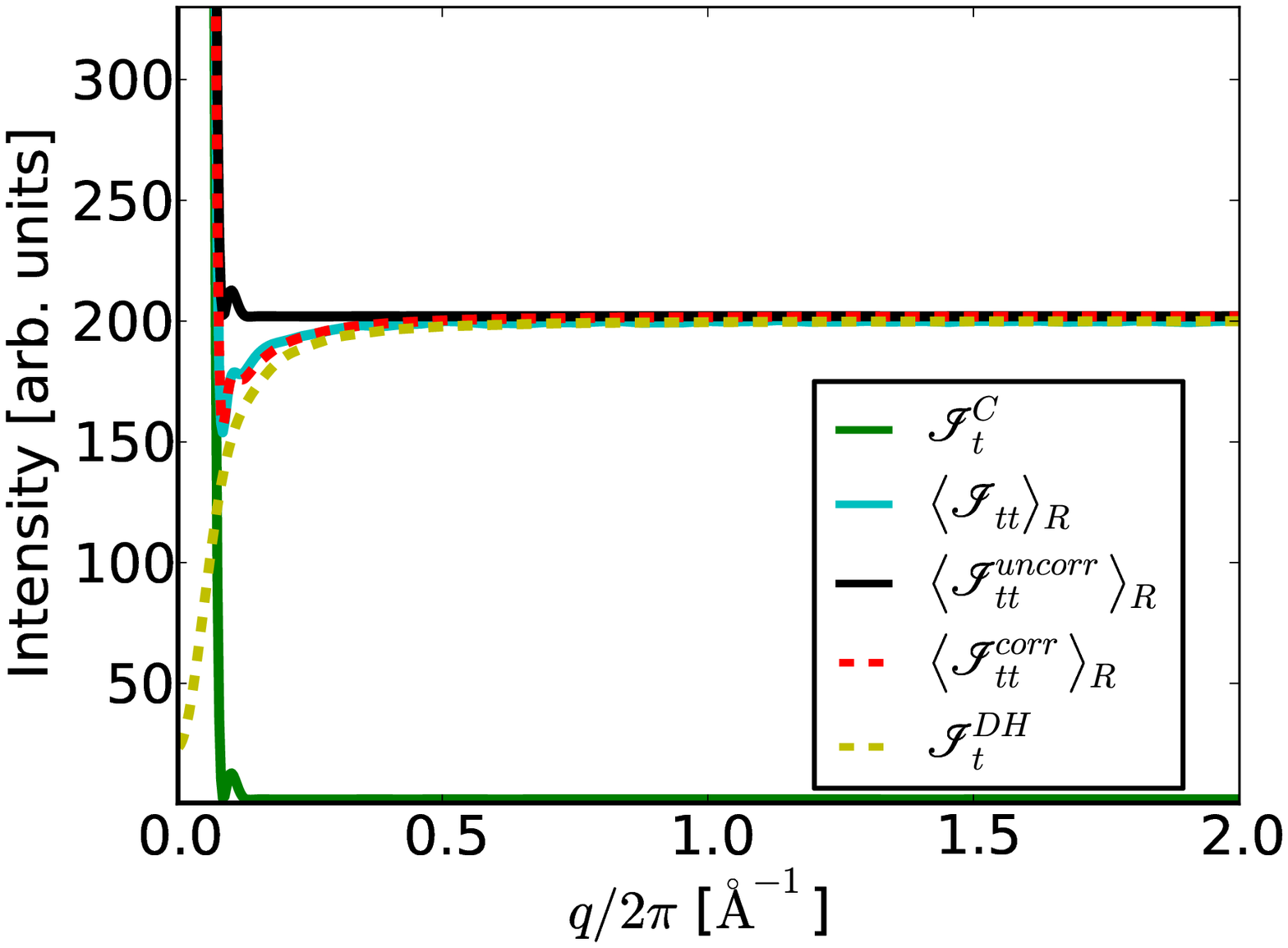} 
\caption{\label{fig:E5h} 
(Color online) Time integrated intensity from diffractive scattering by trapped electrons as a function of the momentum transfer, $q$. The electron temperature was set to 100\,eV, and an external harmonic potential was applied. We show: 
(i) the intensity constructed from average single particle densities $\myotherI_t^C$, (ii) the scattered intensity $ \langle \myotherI_{tt}\rangle _{R}$, (iii) the estimated uncorrelated intensity $ \langle \myotherI_{tt}^{\uncorrsymbol}\rangle _{R}$, (iv) the estimated correlated intensity $ \langle \myotherI_{tt}^{\corrsymbol}\rangle _{R}$, and
(v) the intensity obtained from the theory of homogeneous, infinite-size, weakly coupled plasmas $\myotherI_{t}^{DH}$. 
}
\end{figure}

The electronic system considered here is in thermal equilibrium. The inverse Debye-length:
\begin{eqnarray}
k _D \equiv  \sqrt{ {n_t e^2} \over {\epsilon _0 k_B T} },
\label{eq:DebyeL}
\end{eqnarray}
is $\approx 1\,$\AA\,$^{-1}$\ at the electron temperature $20$\,eV.
In Fig.~\ref{fig:E1h}, we show the time integrated scattered intensity, $ \langle \myotherI_{tt}(\myvect{q}) \rangle _{R}$, along an axis in reciprocal space, averaged over realizations, as compared to the intensity, $\myotherI_t^C(\myvect{q})$, obtained from the average density, and to the intensity, $ \langle \myotherI_{tt}^{\uncorrsymbol}(\myvect{q}) \rangle _{R}$, estimated in the uncorrelated case. The dominant correction (to $\myotherI_t^C(\myvect{q})$) originates from the granularity of electrons, and is given by the total number of the trapped electrons in the system (cf.\ Eq.\ (\ref{eq:trapped_final})).  
At larger momentum transfers, $q > k_D$, this correction is accurate enough
and  $\langle \myotherI_{tt}^{\uncorrsymbol}(\myvect{q}) \rangle _{R}$ agrees well with $ \langle \myotherI_{tt}(\myvect{q}) \rangle _{R}$. However, at lower momentum transfer, it breaks down. The reason for this is that in Eq.~(\ref{eq:trapped_final}) all correlations are neglected a priori, and in our simulated case two-particle correlations are present. They manifest themselves at low $q$.  As expected, their effect decreases at higher electron temperatures, $T=100$\,eV (Fig.~\ref{fig:E5h}). 

For an infinite, homogeneous system, the scattered intensity is connected to the radial distribution function by the following expression \cite{HansenMcD}:
\begin{eqnarray}
\myotherI (\myvect{q}) = N_t \left ( 1 + n_t \int d^3 r \left ( g(\myvect{r}) -1  \right ) 
      e ^ { \timestwopi i \myddot {\myvect{q}} {\myvect{r}} }    \right ),
\label{eq:Itheory}
\end{eqnarray}
where $n_t$ is the homogeneous density of the electrons. In the case of a Debye H{\"u}ckel plasma, the radial distribution function, $g(r)$, is \cite{HansenMcD}:
\begin{eqnarray}
g^{DH}(r) = \exp \left ( - { {k^2_D} \over {4 \pi  n_t \, r}  }  e ^ { - k_D r }   \right ).
\label{eq:gr}
\end{eqnarray}
Apart from the region of $q \lesssim 1/R_0$ , where the finite size effects dominate, at $T=20$\,eV $ \langle \myotherI_{tt}(\myvect{q}) \rangle _{R}$ 
remains in a good agreement with the intensity profile, $\myotherI_{t}^{DH}(\myvect{q})$, obtained from Eqs.\ (\ref{eq:DebyeL}-\ref{eq:Itheory}), evaluated with the initial simulation parameters, $R_0 = 7.1$\,\AA, $n_t = N_t / (4 \pi R_0^3/3)$. A better estimate,  $\langle \myotherI _{tt}^{\corrsymbol}(\myvect{q}) \rangle _R$, can be obtained by evaluating Eq.\ (\ref{itt}) with the pair correlation function, Eq.\ (\ref{eq:gr}). It then applies well in the entire $q$-range.

\begin{figure}
\includegraphics[width = 10cm]{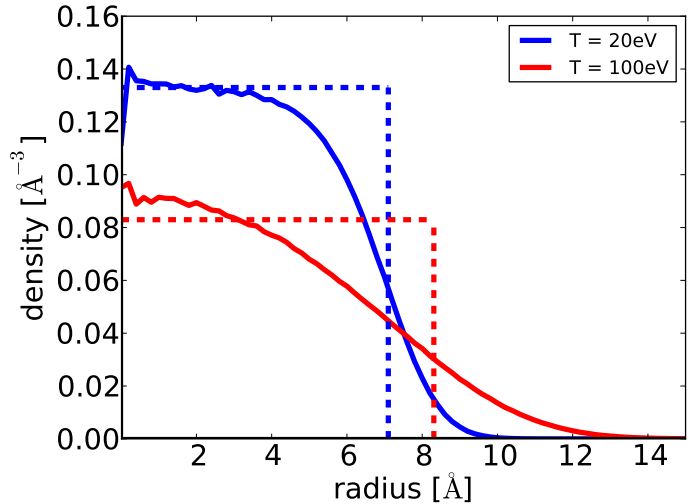} 
\caption{\label{fig:dens} 
(Color online) Average density of an electron system containing 200 electrons (solid lines) at temperatures, 20\,eV and 100\,eV, with an external harmonic potential applied. Dashed lines show the radius and density values, used for evaluating Eq.\ (\ref{itt}) and Eq.\ (\ref{eq:gr}).
}
\end{figure}

At the higher temperature ($T=100\,$eV), the spread of the electron density is wider (Fig.~\ref{fig:dens}). We have then to evaluate
Eq.\ (\ref{itt}) and Eqs.\ (\ref{eq:DebyeL}-\ref{eq:gr}) with parameters that correspond to a homogeneous sphere with the average radius estimated to be $R \approx 8.2$ \AA. Our findings are similar: the intensity fit, $\langle \myotherI _{tt}^{\corrsymbol}(\myvect{q}) \rangle _R$, obtained from Eq.\ (\ref{itt}) is in good agreement with the total signal. 
To sum up, $ \langle \myotherI _{tt}^{\uncorrsymbol}(\myvect{q}) \rangle _{R} $ overestimates the full scattered intensity at $q \lesssim k_D$, however, it still gives an accurate estimate
of the trapped-electron background in the region of high $q$, important for
high resolution imaging. The estimate, $\langle \myotherI _{tt}^{\corrsymbol}(\myvect{q}) \rangle _{R} $ applies well in the entire $q$ region.

\subsection{\label{subsection:bound} Bound and trapped electrons }

\begin{figure}
\includegraphics[width = 10cm]{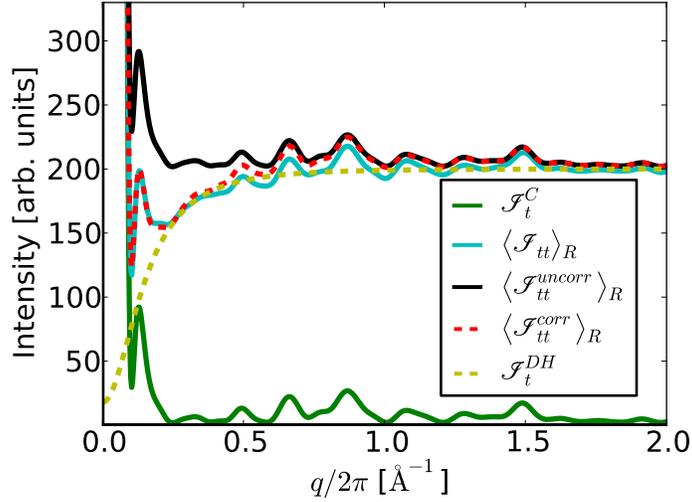} 
\caption{\label{fig:E1c} 
(Color online) Time integrated intensity from diffractive scattering by trapped electrons as a function of the momentum transfer, $q$. The electrons were moving in the field of positive granular ions. The electron temperature was set to 20\,eV. We show (i) the intensity constructed from average single particle densities $\myotherI_t^C$, (ii) the scattered intensity $\langle \myotherI _{tt}\rangle _{R}$, 
(iii) the estimated uncorrelated intensity $ \langle \myotherI^{\uncorrsymbol}_{tt}\rangle _{R}$, (iv) the estimated correlated intensity 
$ \langle \myotherI_{tt}^{\corrsymbol}\rangle _{R}$, and
(v) the intensity obtained from the Debye-H\"uckel theory $\myotherI_t^{DH}$.
}
\end{figure}

At last we discuss the case with contributions from both bound and
trapped electrons. In this simulation electrons were moving in the field of granular ions, therefore we can expect some change in the behavior of the average intensity scattered by the trapped electrons (Fig.~\ref{fig:E1c}) as compared to that scattered by the trapped electrons in
a smooth positive field (Fig.~\ref{fig:E1h}). Discrepancies show up as small
peaks at the $q$ values corresponding to the locations of ion peaks
(Fig.~\ref{fig:I}). These peaks are a consequence of electron-ion correlations resulting in the temporary localization of some electrons near granular ions. Generally, the $q$ dependence of the curves in Fig.~\ref{fig:E1c} is similar to that obtained in the case when a harmonic potential was applied (Fig.~\ref{fig:E1h}). 

\begin{figure}
\includegraphics[width = 10cm]{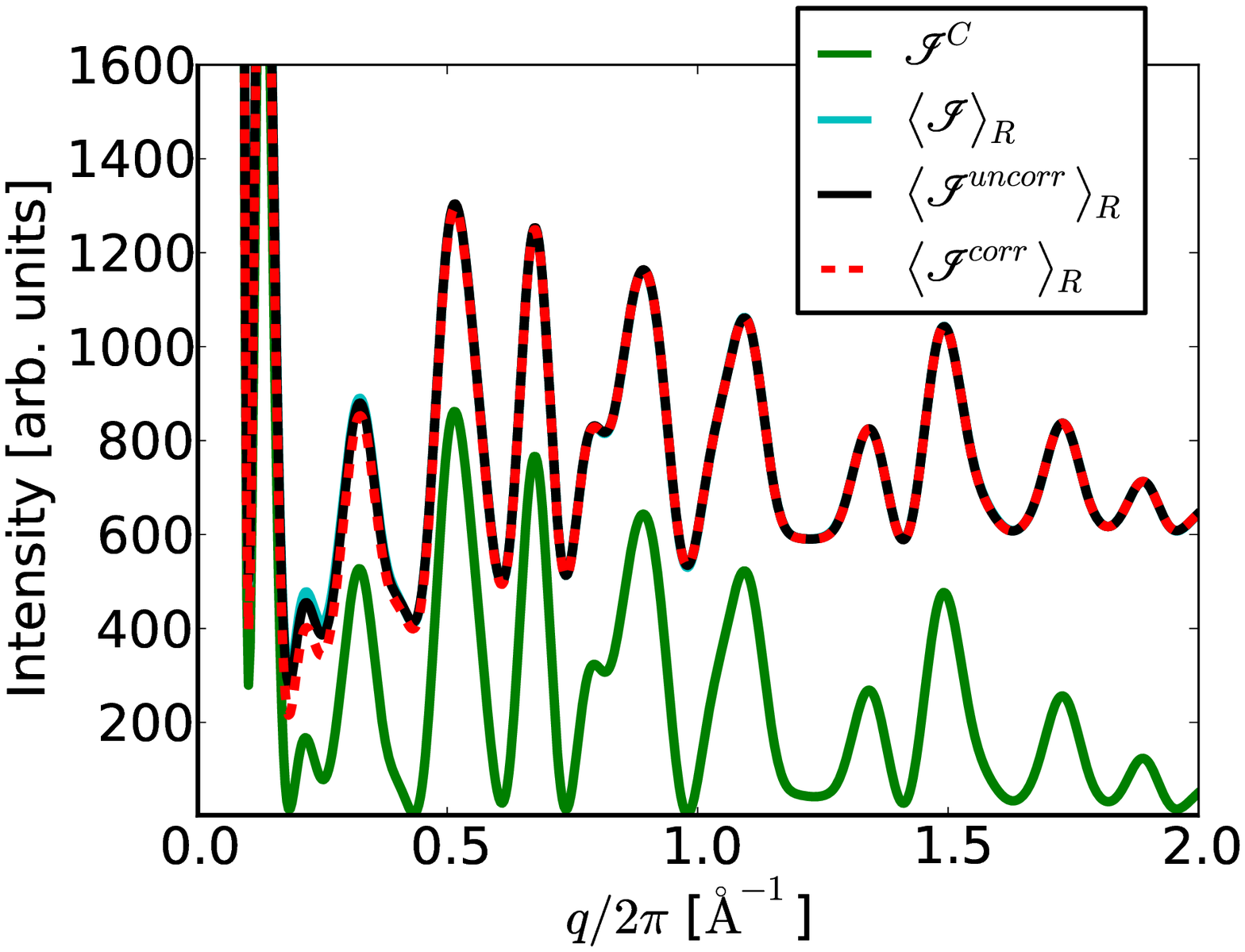} 
\caption{\label{fig:EI} 
(Color online) Time integrated intensity from diffractive scattering by bound and trapped electrons, as a function of the momentum transfer, $q$ along a random axis in reciprocal space. We show (i) the intensity constructed from average single particle densities $\myotherI^C$, (ii) the full scattered intensity $ \langle \myotherI\rangle _{R}$, (iii) the estimated uncorrelated intensity $ \langle \myotherI^{\uncorrsymbol}\rangle _{R}$, and
(iv) the estimated correlated intensity, 
$ \langle \myotherI_{tt}^{\corrsymbol}\rangle _{R}$.}
\end{figure}

The intensities scattered by both bound and trapped electrons are shown in
Fig.\ \ref{fig:EI}. The offset between the intensities $ \langle \myotherI(\myvect{q}) \rangle _{R}$ and $\myotherI^C(\myvect{q})$ at high $q$ is due to the trapped electron contribution, $N_t$, and the inelastic scattering on the bound electrons. For the considered case even the intensity $\langle \myotherI ^{\uncorrsymbol}(\myvect{q}) \rangle _{R}$ that neglects correlations between the trapped electrons, approximates well the full scattering signal. In the calculations presented, $N_e = 0$.

Let us recall here that the structural information is carried by the elastically scattered photons. However, in currently planned experiments on non-periodic single objects, inelastic scattering will also contribute to the measured patterns. Coherent XFEL sources have been reported to produce radiation with a bandwidth of $0.2-1$\,\% \cite{LCLS}. At the photon energy of $E_{ph}=12$\,keV, it corresponds to a bandwidth of $\Delta E_{ph}=20-120$\,eV that is comparable with the Compton broadening. As a consequence, even if detecting exactly at the photon energy $E_{ph}$, one collects the elastic scattering signal of incoming photons and inelastic scattering signal of incoming photons with the initial energy within the bandwidth. Both the elastic and the inelastic scattering then contribute to the measured signal. 

\begin{figure}
(a)\includegraphics[width = 10cm]{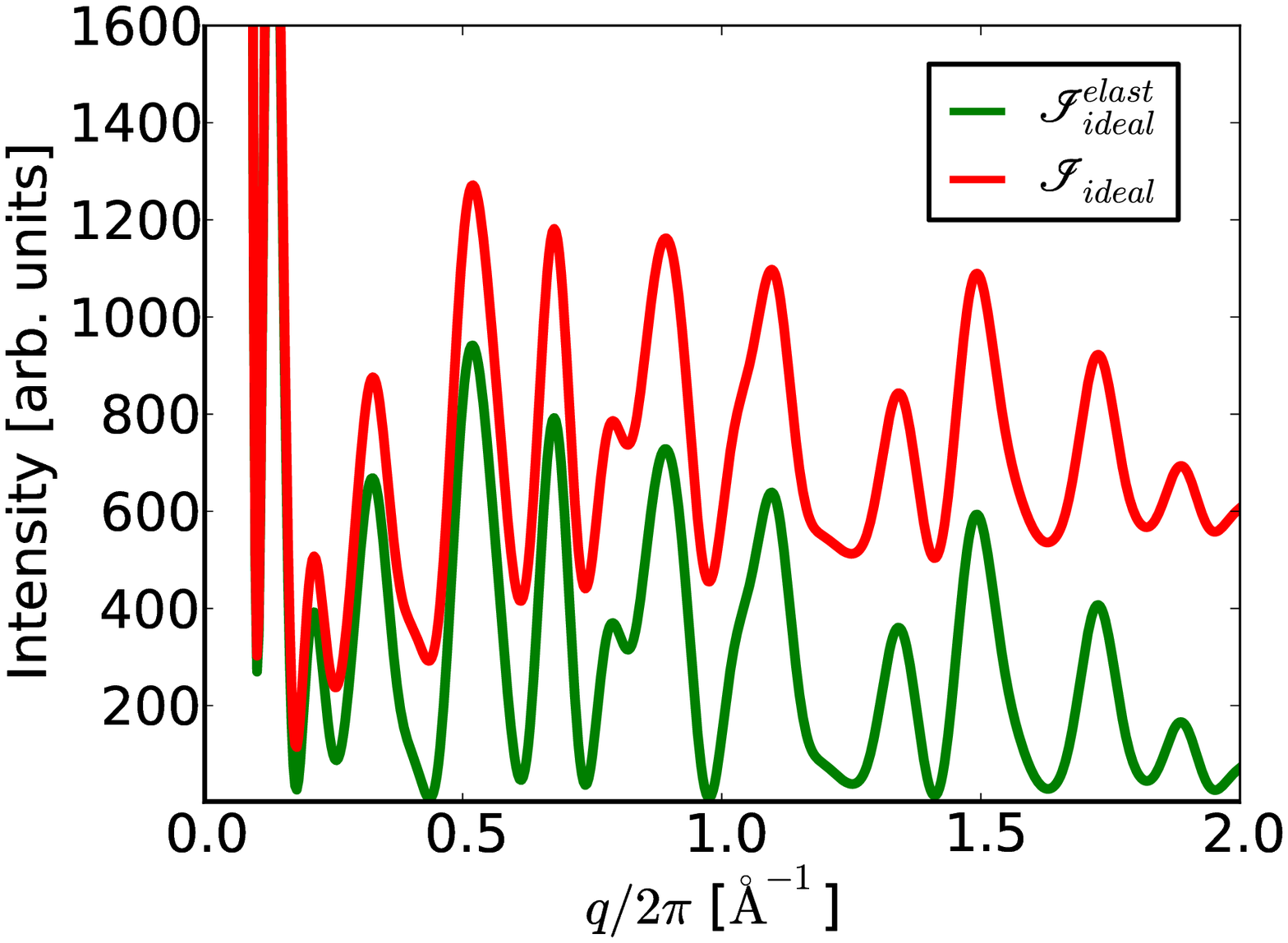} \\
(b)\includegraphics[width = 10cm]{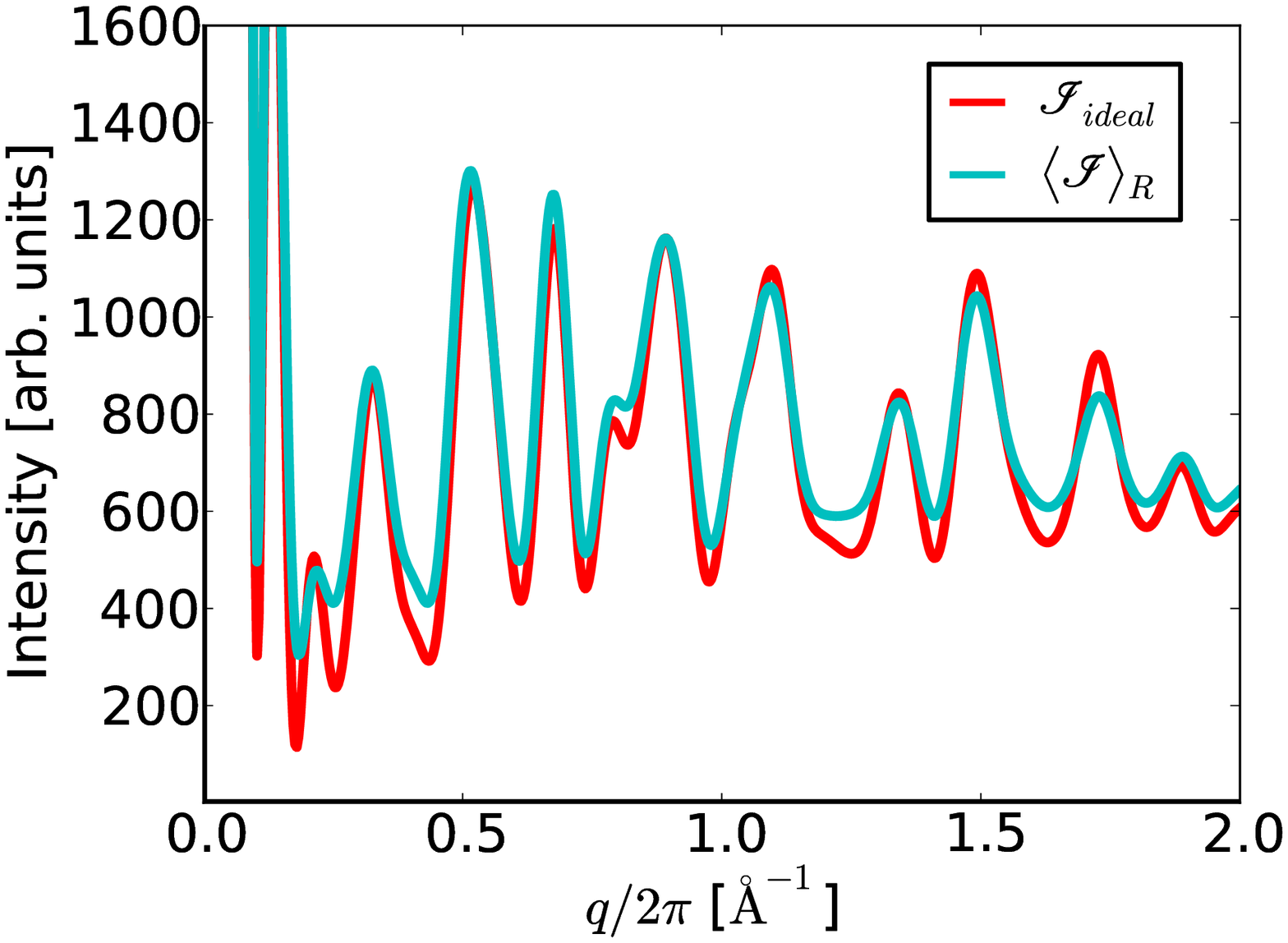} 
\caption{\label{fig:damage}
(Color online) The effect of inelastic scattering and damage on the scattered
intensity. We show: (a) the signal from the 'ideal' (undamaged) sample without inelastic scattering $\myotherI^{elast}_{ideal}$, and with inelastic scattering $\myotherI^{}_{ideal}$, (b) the signal from the 'ideal' (undamaged) sample $\langle \myotherI \rangle _{R}$, and the signal from the damaged sample $ \langle \myotherI \rangle _{R}$, both including the inelastic scattering component.
}
\end{figure}

We showed that inelastic scattering on bound electrons can have a significant impact on the measured intensities:  it contributes to the background that reduces contrast of the recorded image. This effect is even more pronounced at larger momentum transfers ($ \myotherI ^{elast}_{bb,ideal}$  and $ \myotherI ^{}_{bb,ideal}$ in Fig.~\ref{fig:damage}a).

At experiments that have already been done at LCLS, the effect of inelastic
scattering was negligible. When the sample is a nanocrystal
\cite{chapman:Nature2011}, the strong coherent Bragg peaks dominate over inelastic scattering. Inelastic scattering is also negligible at low resolution experiments on single objects \cite{seibert:Nature2011}, when intensity at small scattering vectors is collected. However, the above predictions show that the effect of inelastic scattering should be taken into account, when planning atomic resolution imaging of non-periodic samples.

Finally, we show the effect of damage on the recorded total signal (Fig.~\ref{fig:damage}b). As expected, progressing damage does not change the positions of intensity peaks that correspond to the positions of imaged ions. It only changes the (vertical) positions of intensity minima and maxima, which reduces the image contrast.

\section{\label{section:summary} Summary }

The continuum approach is an efficient tool for the study of the dynamics of large atomic or molecular samples exposed to an intense XFEL pulse.  However, the presently developed continuum models deliver only information on
single-particle densities, and imaging studies require information on
two-particle correlations. 

Here we have studied in detail the effect of two-particle correlations on
CDI scattering patterns, obtained from electron-ion systems under conditions
similar to those expected during XFEL imaging experiments at atomic resolution.
We derived an estimate for the scattered intensities from the single-particle-density continuum model: Eq.\,(\ref{eq:final}), and demonstrated through numerical simulations that it can describe the scattered signal with a good accuracy. Correlation effects manifest themselves only in the region of low $q$, together with the effects of the finite size of the sample. 

Our results have implications for imaging-oriented studies of radiation damage performed with continuum models, as they define the limits of applicability of these models for CDI simulations.

\appendix*
\section{Simulation details}

The simulations performed in this work were based on the molecular dynamics approach. The particles (ions and electrons) were treated as classical, charged point-like particles. 

\begin{itemize}

\item [$\bullet$] { \bf Ionization of atoms }\\
In order to describe the time dependence of the average number of bound electrons per atom, we used an approximate exponential form:
\begin{eqnarray}
N(t) = (N_0-N_{inf}) \, \exp(-wt) + N_{inf},
\label{eq:exp_func}
\end{eqnarray}
where $N(t)$ denotes the number of bound electrons per atom at time $t$. Initially at $t=0$ fs, $N_0=6$. The fitting parameters, $w$ and $N_{inf}$, were adjusted to $N_{inf} = 3.98$ and $w =0.456\,$(fs)$^{-1}$, following the results from Ref.\ \cite{HauRiegePRE2004}. Atomic form factors were calculated with the XATOM package \cite{SanKil2011xatom,xatomcode}.

\item [$\bullet$] { \bf Dynamics of trapped electrons }\\
The motion of the trapped electrons was followed with the Newton equations
solved by the numerical Velocity-Verlet algorithm. The interaction between the
charged particles was described by the regularized Coulomb potential, 
$V(r) = 1/\sqrt{r^2 + r_0^2}$, where the cut-off parameter, $r_0=0.1$\,\AA\,, was used. The simulation timestep was $1$\,as .

Simulating the smooth positive background field for trapped electrons, we applied an external radial harmonic potential of the form, $V(r) = -D\,r^2/2$. Initial electron positions corresponded to those ones within an electron 
system in thermal equilibrium.

Simulating the case with granular ions, we placed the ions at random positions
within a sphere of $30$ \AA\, radius. The ion density was set to $1/15$\,\AA$^{-3}$, which corresponds to a typical atomic density within a protein. The minimal atom-atom distance was kept larger than $1.5$\,\AA. Ionic charge was set to $+2$. Let us note that the static electronic charge was fully neutralized by ions inside a sphere of $7.1$\,\AA\, radius.

\item [$\bullet$] { \bf Calculation of scattered intensities }\\
We recorded the total number of the bound electrons at each atom/ion at a time step. We did not distinguish between bound electrons at different atomic orbitals. Therefore, we approximated the scattering factor of an ion with an average  scattering factor, averaged over different electronic configurations. The scattering factor of a trapped electron was 1.

In our simulation, we considered diffractive scattering only from ions inside the neutralizing $7.1$\,\AA-radius sphere (corresponding to the imaged net-neutral electron-ion sample), and not from the ions in the positively charged outer shell.

\end{itemize}

\begin{acknowledgments}
The authors would like to thank G. Dixit, C. Fortmann, S. P. Hau-Riege, A. V. Martin for illuminating discussions.
\end{acknowledgments}


\providecommand{\noopsort}[1]{}\providecommand{\singleletter}[1]{#1}%

\end{document}